\documentclass[aps,prd,twocolumn,nofootinbib,superscriptaddress]{revtex4}

	\usepackage{amssymb}
	\usepackage{amsmath}
	\usepackage{mathtools}
	\usepackage{graphicx}
	\usepackage{subfigure}

\newcommand{\dint}{\int\!\!\!\!\int}

\begin{document}

\title[Galaxy clusters and dark energy]
{Galaxy clusters and structure formation in quintessence versus phantom
dark energy universe}

\author{Zacharias Roupas}\email{roupas@inp.demokritos.gr}
\affiliation{Institute of Nuclear and Particle Physics, N.C.S.R. Demokritos,
GR-15310 Athens, Greece} 
\author{Minos Axenides}\email{axenides@inp.demokritos.gr}
\affiliation{Institute of Nuclear and Particle Physics, N.C.S.R. Demokritos,
GR-15310 Athens, Greece} 
\author{George Georgiou}\email{georgiou@inp.demokritos.gr}
\affiliation{Institute of Nuclear and Particle Physics, N.C.S.R. Demokritos,
GR-15310 Athens, Greece} 
\author{Emmanuel N. Saridakis}\email{Emmanuel_Saridakis@baylor.edu}
\affiliation{Physics Division, National Technical University of Athens,
15780 Zografou Campus,  Athens, Greece}
\affiliation{Instituto de F\'{\i}sica, Pontificia Universidad  Cat\'olica
de Valpara\'{\i}so, Casilla 4950, Valpara\'{\i}so, Chile}

\begin{abstract}
	The self-gravitating gas in the Newtonian limit is studied in the
presence of dark energy with a linear and constant equation of state. Entropy
extremization associates to the isothermal Boltzmann distribution an effective
density that includes `dark energy particles', which either strengthen or weaken
mutual gravitational attraction, in case of quintessence or
phantom dark energy, respectively, that satisfy a linear equation of state. Stability is studied
for microcanonical (fixed energy) and canonical (fixed temperature)
ensembles. Compared to the previously studied cosmological constant case, 
in the present work it is found that quintessence increases,
while phantom dark energy decreases the instability domain under
gravitational collapse. 
Thus, structures are more easily formed in a quintessence rather than in a
phantom dominated Universe. 
Assuming that galaxy clusters are spherical,
nearly isothermal and in hydrostatic equilibrium we find that dark energy
with a linear and constant equation of state,
for fixed radius, mass and temperature, steepens their total density profile!
In case of a cosmological constant, this effect accounts for a $1.5\%$
increase
in the density contrast, that is the center to edge density ratio of the
cluster. We also propose a
method to constrain phantom dark energy. 
\end{abstract}

\maketitle

\section{Introduction}

\indent Dark energy is considered a main component of our universe, 
since its existence relies on convincing observational data \cite{Hinshaw:2012aka,Amanullah:2010vv,Blake:2011en}. However, its nature still remains a mystery for physics. Three main candidates are quintessence
\cite{Ratra:1987rm,Wetterich:1987fm,Liddle:1998xm,Basilakos:2003hx,
Guo:2006ab,Dutta:2008qn,Dutta:2009yb}, the cosmological constant and phantom dark energy
\cite{Caldwell:1999ew,Caldwell:2003vq,Nojiri:2003vn,Onemli:2004mb,
Saridakis:2008fy,Saridakis:2009pj}. 

In this expanding universe, driven by dark energy, structures form by the gravitational instability 
of the self-gravitating gas, that causes a density perturbation to decouple from expansion and collapse. In the matter dominated era, the velocity of gas' constituents 
is low and the Newtonian limit is appropriate. 
The thermodynamic stability of a self-gravitating gas in the Newtonian
limit is a very old subject
\cite{Antonov:1962,Bell:1968,Padmanabhan:1989,Padmanabhan:1990,
Chavanis:2001hd,deVega:2001zk,deVega:2001zj,Katz:2003} (the
relativistic case has recently earned a lot of attention
\cite{Sorkin:1981,Chavanis:2001ah,Chavanis:2007kn,Gao:2011hh,Roupas:2013nt,Roupas:2013kaa,
Anastopoulos:2013xdk,Green:2013ica,Schiffrin:2013zta,Fang:2013oka}). 
Hence, this subject has not only a pure theoretical interest, 
but also an additional cosmological
motivation. Its direct application to large
structures in the universe, such as the galaxy clusters, provides
information on their formation and  evolution. Furthermore, one can extend
\cite{deVega:2003hm,deVega:2004vv} this analysis and 
investigate the effects of a cosmological constant on the stability 
of the self-gravitating gas
\cite{Axenides:2012bf,Axenides:2013hrq,Axenides:2013hba}.

In the present work we extend these latter studies 
\cite{Axenides:2012bf,Axenides:2013hrq,Axenides:2013hba} in order to
examine the effects of dark energy, satisfying a linear equation of state, on the stability of
isothermal
spheres and galaxy clusters. In particular, we would like to see 
how the galaxy clusters and especially their 
more massive component, namely the dark matter haloes, are affected by the
presence of a quintessence or a phantom dark energy, 
parametrized by a linear equation of state. As we will see, compared to the
simple cosmological constant $\Lambda$ case,
apart from increasing the repulsion due to the negative pressure, phantom
dark energy in the Newtonian description 
introduces effective  ``dark energy particles''
that weaken gravitational attraction. On the other hand, quintessence has
the inverse effects, that is it decreases repulsion
due to pressure and introduces dark energy particles that strengthen mutual
gravitational attraction compared to the $\Lambda$ case. 
Thus, for both reasons, phantom dark energy decreases the 
(under gravitational collapse) instability domain and 
quintessential dark energy increases the instability domain with respect to
the $\Lambda$ case. Large-scale structures are more
difficult to be formed in a phantom universe rather than in a quintessence
one. This is one main result of our analysis.

The potential effect of a dark-energy component on the formation
of galaxy clusters has been inspected recently (see
\cite{Mota:2004pa,Basilakos:2009mz} and References therein).
As we will see, the dark-energy sector does indeed
have effects on the density profile of galaxy clusters and
mainly their most massive component, the dark matter haloes. 
We find that the effects are 
in principle detectable, and more interestingly that the density profile
steepens, in
contrast with naive expectation. Additionally, one can use these results
the other way around, and impose constraints on the dark-energy
equation-of state from galaxy-clusters observations. 

The paper is organized as follows. In section \ref{sec:instab} we
present the thermodynamics of self-gravitating gas in the presence of dark
energy and we study its effect on stability, studying
hydrostatic equilibrium, calculating the entropy extrema and
 performing the stability analysis. In section
\ref{sec:clusters} we study the galaxy clusters,
investigating the effect of dark energy on the clusters' density profile and
proposing a method to constrain the dark
energy equation of state, based on clusters observations. Finally, section
\ref{Conclusions} is devoted to the conclusions.

\section{Thermodynamic instabilities and dark energy}\label{sec:instab}

In order to investigate the dark energy effects on the stability of a
self-gravitating gas, we
will be based on our previous studies
\cite{Axenides:2012bf,Axenides:2013hrq,Axenides:2013hba}. 
We assume that the dark
energy sector is described by a perfect fluid of energy density
$\rho_X$ and pressure $p_X $, while its equation-of-state parameter is
defined as 
\begin{equation}\label{eq:eos_de}
	w\equiv\frac{p_X}{c^2\rho_X},
\end{equation} 
where for clarity we keep the light speed $c$ in the equations. In the
following we restrict ourselves to the observationally favored case $w <
-\frac{1}{3}$, although this is not necessary. In this work we assume that $w=const$, 
as we want to consider the simplest possible setup, 
in order to understand the basic effects of dark energy.
The extension to the full time-varying $w$ and/or time-varying
cosmological constant \cite{Shapiro:2000dz,Polyakov:2009nq,Basilakos:2010rs,Sola:2013gha,Basilakos:2013xpa}, 
as well as the incorporation
of possible dark-energy dark-matter interactions 
\cite{Wetterich:1994bg,Billyard:2000bh,Boehmer:2008av,Chen:2008ft}
and the corresponding complicated analysis,
will follow in a subsequent work.

Let us now derive the modified Poisson equation with a dark energy component
in the Newtonian limit. 
Denoting by $\rho$ and $p$ the gas' energy density and pressure,
respectively, 
we can write down the total 
 energy-momentum tensor as
\begin{equation}
\label{eq:EMtens}
	T^\mu_\nu = \left[(p+p_X) + (\rho+\rho_X)
c^2\right]g_{\alpha \nu} \frac{dx^\mu}{ds}\frac{dx^\alpha}{ds} - (p+p_X)
\delta^\mu_\nu,
\end{equation}
where $g_{\alpha \nu}$ is the spacetime metric with sign $[+,-,-,-]$, $x^\mu$
the spacetime coordinate of one fluid element and $s$ the proper length. In
the non-relativistic limit, and assuming equilibrium (that is $d\vec{x}/ds
= 0$) it
becomes:
\begin{equation}
	T^\mu_\nu \simeq [\rho + (1+w)\rho_X]c^2\,
\delta^\mu_0 \delta^0_\nu - w\rho_X c^2\,\delta^\mu_\nu.
\end{equation}
Defining  the gravitational potential  $\phi$ as usual through 
\[
	\frac{d^2\vec{x}}{dt^2} = -\vec{\nabla} \phi,
\]
we can calculate that for slowly moving particles $\Gamma^i_{00} \approx d^2
x^i/(c^2 dt^2)=\partial^i\phi/c^2$, and thus that in the 
  static weak field (Newtonian) limit we have
$
	R^0_0 =      \frac{1}{c^2}\nabla^2\phi.
$
Inserting these Newtonian-limit expressions in the  time-time component of
the Einstein's equations
\begin{equation}
\label{eq:Ein_2}
	R^\mu_\nu = \frac{8 \pi G}{c^4} T^\mu_\nu - \frac{4\pi G}{c^4} T
\delta^\mu_\nu,
\end{equation}
we finally obtain
\begin{equation}\label{eq:Poiss_DE}
	\nabla^2 \phi(r) = 4\pi G\rho + 4\pi G (1+3w)\rho_X.
\end{equation}
This is the modified Poisson equation that determines the
gravitational potential in the Newtonian limit. \\

\subsection{Hydrostatic equilibrium}\label{sec:HE}

It will be instructive to study the hydrostatic equilibrium of the self-gravitating
gas in presence of dark energy. 
We start with the relativistic
equation of hydrostatic equilibrium, known as Tolman-Oppenheimer-Volkof (TOV)
equation \cite{Tolman:1939,Oppenheimer:1939}, which however we need to derive
in the presence of the dark energy component. As we show in detail in 
Appendix \ref{app:A},  in the static, spherically symmetric case, 
 the Einstein's equations reduce to two equations, namely
\begin{widetext}
\begin{align} \label{eq:TOV1}	
&
\frac{dp}{dr} = -\left[\frac{p}{c^2} + \rho +
(1+w)\rho_X\right] 
\left[
\frac{G\mathcal{M}(r)}{r^2} + 4\pi G \frac{p}{c^2}r + \frac{4\pi
G}{3}\rho_X r(1+3w)\right]
  \left( 1 - \frac{2G\mathcal{M}(r)}{rc^2} - \frac{8\pi
G}{3c^2}\rho_X r^2\right)^{-1}		
\\
\label{eq:mprime} 
& \frac{d\mathcal{M}(r)}{dr} = 4\pi\rho r^2,
\end{align}
\end{widetext}
where $\rho(r)$ and $p(r)$  are respectively the total mass-energy density
and pressure at point $r$, and
$\mathcal{M}(r)$ the total mass-energy contained inside $r$. The
first equation
(\ref{eq:TOV1}) is the TOV equation  in the presence of a dark-energy 
component. In the Newtonian limit, that is for $c
\rightarrow \infty$, we obtain
\begin{equation}\label{eq:Neq}
	\frac{d p}{dr} = -\left[\rho + (1+w)\rho_X\right]\left[
G\frac{\mathcal{M}(r)}{r^2} + \frac{4\pi G}{3}\rho_X r (1+3w)
\right],
\end{equation}
where $\rho$ now is the density of matter. This is the equation of
hydrostatic equilibrium in the presence of dark energy in the Newtonian limit.

We observe, that gravity is now exerted on effective matter with density
\begin{equation}\label{eq:rhoeff_def}
	\rho_{\mbox{\footnotesize eff}} = \rho + (1+w)\rho_X.
\end{equation}
This definition for the effective matter density is
also inferred by the momentum component of the energy-momentum tensor in the
Newtonian
limit, namely
\begin{equation}\label{eq:momentum}
	 T^{0i} = [\rho + (1+w)\rho_X]\frac{dx^i}{dt},
\end{equation}
as it straightforwardly arises from (\ref{eq:EMtens}).

Note that looking at the Poisson equation (\ref{eq:Poiss_DE}), a
naive guess would be to define the effective density as
$\rho+(1+3w)\rho_X$ instead of (\ref{eq:rhoeff_def}). This would be wrong.
Dark energy introduces
an attractive part  coming from $\rho_X$ and a repulsive one coming from
the negative pressure with three components. In case of a
cosmological constant ($w=-1$) the attractive part is completely
counterbalanced by the one pressure component, as is evident by the momentum
(\ref{eq:momentum}), leaving only a term $-2\rho_X$ in the Poisson equation,
without the need of introducing any kind of new matter. However, we see that in the general
dark-energy case, apart from the repulsive gravity due to the pressure, in
the Newtonian limit  we have the effective appearance of additional ``matter
particles'' that gravitate normally in the
quintessence case ($\rho_{\mbox{\footnotesize eff}} > \rho$) or that 
tend to gravitationally neutralize normal matter
in the phantom case ($\rho_{\mbox{\footnotesize eff}} < \rho$). 
 
Let us determine the density distribution for which the equation of
hydrostatic equilibrium (\ref{eq:Neq}) 
leads to the modified Poisson equation (\ref{eq:Poiss_DE}). For an isothermal
distribution $T=const.$, the velocity distribution of the gas particles
should be a a Maxwellian:
\begin{equation}
\label{eq:MB_iso}
	f(r,\upsilon) = \left(\frac{m}{2\pi kT}\right)^\frac{3}{2}
\frac{\rho_{\mbox{\footnotesize eff}}(r)}{m} e^{-\frac{m}{kT}\upsilon^2/2},
\end{equation}
with $m$ their masses, $v$ their velocities, and $k$ the Boltzmann constant.
Respectively, the pressure writes as
\begin{equation}
\label{eq:p_ef}
	p(r) \equiv \int f\frac{1}{3}m\upsilon^2 d^3\upsilon =
\rho_{\mbox{\footnotesize eff}}(r)\frac{kT}{m}.
\end{equation}
Thus, in order to get (\ref{eq:Poiss_DE}) from (\ref{eq:Neq}) we should have
\begin{equation}
\label{eq:rhoeff_dis} 
\rho_{\mbox{\footnotesize eff}} = \rho_{0,\mbox{\footnotesize eff}}
e^{-\frac{m}{kT} [\phi - \phi(0)]},
\end{equation}
which is just the Boltzmann distribution for $\rho_{\mbox{\footnotesize
eff}}$. 
Inserting this to (\ref{eq:rhoeff_def}) we acquire
\begin{equation}
\label{eq:rho_dis} 	
\rho = [\rho_0 + (1+w)\rho_X]
e^{-\frac{m}{kT} [\phi - \phi(0)]} - (1+w)\rho_X.
\end{equation}
Finally,
substituting
expressions (\ref{eq:p_ef}), (\ref{eq:rhoeff_dis}) and (\ref{eq:rho_dis}) into
equation of hydrostatic equilibrium
(\ref{eq:Neq}), we finally obtain
\begin{equation}
\label{eq:Pois_SPH} 
\frac{1}{r^2}\frac{d}{dr}\left(
r^2\frac{d\phi}{dr}\right) = 4\pi G\rho + 4\pi G \rho_X (1+3w),
\end{equation}
that is the spherically symmetric version of the Poisson
equation (\ref{eq:Poiss_DE}).

The fact that the effective density as defined in (\ref{eq:rhoeff_def}) 
does obey Boltzmann distribution, reassures us that it is the correct choice.
We stress that the correct definition of the effective matter
density is crucial, since it affects the calculation of the potential energy,
but most importantly because it is the one measured in indirect mass
observations (for instance in gravitational lensing measurements).

\subsection{Entropy extremum}\label{sec:EE}

Let us prove that the distributions (\ref{eq:MB_iso}), (\ref{eq:rhoeff_dis}) 
extremize the entropy and thus, that they describe thermodynamic equilibria. 
Let the self-gravitating gas be bounded by spherical walls. 
This condition is needed for the entropy to have an extremum. 
Equivalently, only under this condition can hydrostatic equilibrium exist for
finite mass.
Such a configuration is called an ``isothermal sphere''
\cite{Bell:1968,Binney:1987}.

Let an isothermal sphere have radius $R$, and let 
\begin{equation}
\label{eq:S_bolt}
	S = -k\int f(\vec{r},\vec{\upsilon}) \log f(\vec{r},\vec{\upsilon})
d^3\vec{r}d^3\vec{\upsilon},
\end{equation}
be the Boltzmann entropy, where the distribution $f(\vec{r},\vec{\upsilon})$ 
provides the number of
effective particles that are inside the cube $d^3\vec{r}$ at $\vec{r}$, with
velocities from $\vec{\upsilon}$ to $\vec{\upsilon} + d\vec{\upsilon}$.
Thus, we have
\begin{equation}
\label{eq:rhoeff_int}
	\rho_{\mbox{\footnotesize eff}}(\vec{r}) = m \int
f(\vec{r},\vec{\upsilon}) d^3\upsilon,
\end{equation}
and the total effective mass is 
\begin{equation}\label{eq:Meff}
	M_{\mbox{\footnotesize eff}} = m \int  f d^6\tau,
\end{equation}
with $d^6\tau = d^3\vec{r}d^3\vec{\upsilon}$.
In order to calculate the distribution $f$ that extremizes the entropy $S$,
we have to calculate the variation of (\ref{eq:S_bolt}) in terms of $f$. 

The Poisson equation (\ref{eq:Poiss_DE})
can be written as
\begin{equation}
	\nabla^2 \phi(r) = 4\pi G\rho_{\mbox{\footnotesize eff}} + 8\pi G w \rho_X.
\end{equation}
We see that the effective particles interact
mutually with Newtonian gravity and moreover interact with some repulsive
potential. Hence, we can define the effective potentials
\begin{align}
\label{eq:PhiNef}
&	\phi_{N(\mbox{\footnotesize eff})} = -G \int
\frac{\rho_{\mbox{\footnotesize eff}}(\vec{r}\,')}{|
\vec{r}-\vec{r}\,'|}d^3\vec{r}\,'
\\
\label{eq:PhiLef}
&	\phi_{X(\mbox{\footnotesize eff})} = \frac{4\pi G}{3} w \rho_X r^2,
\end{align}
with the total potential being
\begin{equation}\label{eq:phiTOT}
	\phi = \phi_{N(\mbox{\footnotesize eff})} + \phi_{X(\mbox{\footnotesize eff})}
\end{equation}
and therefore the total potential energy takes the simple form
\begin{equation}
\label{eq:U_eff}
	U = \frac{1}{2}\int \rho_{\mbox{\footnotesize eff}}
\phi_{N(\mbox{\footnotesize eff})} d^3\vec{r} + \int
\rho_{\mbox{\footnotesize eff}} \phi_{X(\mbox{\footnotesize eff})}
d^3\vec{r}.
\end{equation}
Inserting (\ref{eq:rhoeff_int}) and (\ref{eq:PhiNef}) into
(\ref{eq:U_eff}), we straightforwardly find
\begin{align}
\nonumber	U = &-\frac{G}{2}\dint  m^2f(\vec{r},\vec{\upsilon})
\frac{f(\vec{r}\,',\vec{\upsilon}\,')}{|\vec{r}-\vec{r}\,'|}d^6\tau\,'
d^6\tau 
\\
\label{eq:U_fin} 
& + \int  mf \phi_{X(\mbox{\footnotesize eff})} d^6\tau.  
\end{align}

We want to extremize the entropy with constant energy and mass. Using the
Lagrange's multipliers method, the following variation condition should be
satisfied to first order:
\begin{equation}\label{eq:Sfirst}
	\delta S/k - \beta\delta E + \alpha \delta M_{\mbox{\footnotesize
eff}} = 0,
\end{equation}
where $\beta$, $\alpha$ are two yet undetermined Lagrange multipliers and $k$
the Boltzmann's constant. 
Inserting $\delta S$ from (\ref{eq:S_bolt}) and $\delta M_{\mbox{\footnotesize
eff}}$ from (\ref{eq:Meff}), as well as calculating from 
(\ref{eq:U_fin}) that
$
	\delta E =m \int  \delta f \left(\upsilon^2/2 + \phi - C
\right)d^6\tau,
$
with $C = 2\pi G (1+w)\rho_X R^2$, we finally acquire from (\ref{eq:Sfirst})
that:
\begin{equation}
 	 \log f + 1 + m \beta \left( \frac{\upsilon^2}{2} + \phi - C\right) -
m\alpha = 0.
\end{equation}
This finally gives
\begin{equation}
\label{fcalculated}
	f(r,\upsilon) = A e^{-\beta m\left[\frac{1}{2}\upsilon^2 +
\phi(r)\right]},
\end{equation}
where $A = e^{m\alpha  - 1 + m\beta C}$. From this expression we derive
the average kinetic energy per particle
\[
	\frac{\int f\frac{1}{2}m\upsilon^2d^3\vec{\upsilon}}{\int f
d^3\vec{\upsilon}} = 
	\frac{\int e^{-m\beta
\left(\frac{1}{2}\upsilon^2\right)}\frac{1}{2}m\upsilon^2d^3\vec{\upsilon}}{
\int e^{-m\beta \left(\frac{1}{2}\upsilon^2\right)} d^3\vec{\upsilon}} 
	= \frac{3}{2\beta},
\]
and therefore we see that $\beta$ should be interpreted as the inverse
temperature 
\begin{equation}
	\beta = \frac{1}{kT}.
\end{equation}
	
Finally, inserting the calculated $f(r,\upsilon)$ from (\ref{fcalculated})
into equation (\ref{eq:rhoeff_int}) we obtain
\[
	\rho + (1+w)\rho_X =   \left(\frac{2\pi kT}{m}\right)^{\frac{3}{2}}
mA
e^{-\frac{m}{kT}\phi(r)},
\]
which for 
\[
	mA = \left(\frac{m}{2\pi kT}\right)^{\frac{3}{2}} [\rho_0 +
(1+w)\rho_X]e^{\frac{m}{kT} \phi(0)} 
\]
leads to equation (\ref{eq:rho_dis})
\begin{equation}\label{eq:rho_DIS}
	\rho(r) = [\rho_0 + (1+w)\rho_X]e^{-\frac{m}{kT} [\phi(r) -
\phi(0)]} - (1+w)\rho_X 
\end{equation}
and to the distribution (\ref{eq:MB_iso})
\[
	f(r,\upsilon) = \left(\frac{m}{2\pi kT}\right)^\frac{3}{2}
\frac{\rho_{\mbox{\footnotesize eff}}(r)}{m} e^{-\frac{m}{kT}\upsilon^2/2},
\]
as desired. 

In summary, we see that the hydrostatic equilibrium of a
self-gravitating gas, in the presence of dark energy, for an isothermal
distribution corresponds to a state of
entropy extremum, with an effective density given by (\ref{eq:rhoeff_dis})
that obeys the Boltzmann distribution. 
However, the stability of this state depends on whether
the extremum is maximum or not.

\subsection{Energy, temperature and stability}\label{sec:ETstab}

In order to determine the  type of the entropy extremum, and thus to deduce
whether we have stability or not, we have to examine the sign of the second
variation of entropy. Fortunately, due to a theorem of Poincar\'e
\cite{Poincare}, and its subsequent refinements by Katz
\cite{Katz:1978,Katz:2003},
one does not always have to calculate the second variation of entropy or free
energy. In particular, for  fixed energy $E$ and mass $M$, that is in the
case of  microcanonical ensemble, an instability sets in at the equilibrium
point where there is a vertical tangent on the diagram of equilibria $T(E)$.
On the other hand, for fixed temperature $T$
and mass $M$, that is in the case of  canonical ensemble, an instability
sets in at the equilibrium point where there is a vertical tangent on the
diagram of equilibria $E(T)$. For an isothermal sphere of a self-gravitating
gas, the instability in the microcanonical ensemble is called gravothermal
catastrophe \cite{Antonov:1962,Bell:1968,Padmanabhan:1989,Axenides:2012bf,
Axenides:2013hrq}, while the instability in the canonical ensemble is called
isothermal collapse \cite{Axenides:2013hrq,Chavanis:2001hd,
Axenides:2013hba}. 

In practice, in a 
series-of-equilibria diagram of the energy versus any variable, an extremum
is a turning point of stability. This implies that if one branch of the
diagram up to the turning point is known to be stable, the branch beyond the
turning point will be unstable, and vice versa. In addition, since we have an
extremum, there do not exist
equilibria at all above this extremum in the case where it is a global
maximum, or beneath this extremum in the case where it is a global minimum.
Therefore, we conclude that the system exhibits  a gravothermal
catastrophe  at the turning point in a series-of-equilibria diagram of
$E$ versus any variable, and similarly it exhibits an isothermal collapse at
the turning point in a series-of-equilibria diagram of $T$ versus any
variable.

In conclusion, instead of having to perform the complicated calculation of
the second variation of entropy, we just need to  calculate the energy and
temperature of the isothermal sphere in the presence of dark energy, and
draw the corresponding diagrams of series of equilibria. 

Substituting the density distribution (\ref{eq:rho_DIS}) into the
Poisson equation (\ref{eq:Pois_SPH}), we obtain the Emden equation
\cite{Binney:1987},
modified with the additional contribution of the dark-energy
component, namely
\begin{align}
\nonumber 	
 \frac{1}{r^2}\frac{d}{dr}\left( r^2\frac{d\phi}{dr}\right) = \;
 & 4\pi
G[\rho_0 + (1+w)\rho_X]e^{-\frac{m}{kT} [\phi(r) - \phi(0)]} \\
&
\label{eq:Emden_Dim}
+ 8\pi G
w\rho_X ,
\end{align}
where $\rho_0$ is the density of matter at the center of the sphere. 
Defining the dimensionless variables
\begin{equation}
\label{eq:variables}
	x = r\sqrt{4\pi G \rho_0 \frac{m}{kT}} \; ,\; y = \frac{m}{kT} [\phi
- \phi(0)] \; , \; \lambda = \frac{2\rho_X}{\rho_0},
\end{equation}
the modified Emden equation (\ref{eq:Emden_Dim}) becomes:
\begin{equation}\label{eq:Emden}
	\frac{1}{x^2}\frac{d}{dx}\left( x^2\frac{d}{dx}y\right) = \left( 1 +
\frac{1+w}{2}\lambda\right) e^{-y} + w\lambda,
\end{equation}
with initial conditions $y(0) = y'(0) = 0$. 
Hence, for given values of $w$ and $\rho_X$, an
equilibrium configuration is
completely determined by the values of mass $M$ and radius $R$ that
correspond to a temperature $T$ and energy $E$. In the following we
desire to numerically generate a series of equilibria $T(E)$ for fixed
mass $M$ in order to study stability, which is not  straightforward since
there are some complications.

Let $z$ be the value of $x$ at $R$:
\begin{equation}\label{eq:z_var}
	z = R\sqrt{4\pi G \rho_0 \frac{m}{kT}}.
\end{equation}
To obtain a solution of equation (\ref{eq:Emden}) one has to specify the
couple $(z,\lambda)$. As we noted, we do not want just a solution, but a
consistent series of solutions (equilibria). Assuming some fixed value of
$\rho_X$, then fixing $\lambda$ and solving for various $z$ would generate a
series with different mass at each equilibrium. The reason is that fixing
$\lambda$ corresponds to fixing $\rho_0$. Therefore, in order to have
different equilibria for various radii $R$ and hence $z$, these equilibria
should have different mass. The same holds if one keeps $z$ constant and vary
$\lambda$. The deeper reason for this difficulty in determining a
consistent series of equilibria, that does not exist without
dark energy, is that dark energy introduces a mass
scale 
\begin{equation}
	M_X = \frac{4}{3}\pi R^3 \rho_X
\end{equation} 
to the system. Thus, we introduce a dimensionless mass:
\begin{equation}\label{eq:M_dless}
	\mu = \frac{M}{2M_X} = \frac{<\rho>}{2\rho_X}.
\end{equation}
Based on earlier works \cite{Axenides:2012bf,Axenides:2013hrq,Roupas:2013nt}
we construct a computer code that can solve equation (\ref{eq:Emden}) for
various values $(z,\lambda)$ and choose these solutions that correspond to a
fixed (up to some tolerance determined by the user) value of $\mu$. In this
way, we can generate consistent series of equilibria corresponding to  the
same mass. Performing the calculation for various $\mu$ we can generate the
series for various values of $\rho_X$. 

In order to proceed to this numerical elaboration, we define 
a dimensionless energy 
\begin{equation}\label{eq:E}
	\mathcal{E} = \frac{ER}{GM^2},
\end{equation}
and a dimensionless inverse temperature 
\begin{equation}\label{eq:B}
	\mathcal{B} = \frac{GMm}{RkT}.
\end{equation}
In order to calculate $\mathcal{B}$ we integrate equation (\ref{eq:Pois_SPH}),
using also the dimensionless variables (\ref{eq:variables}), obtaining 
\begin{equation}\label{eq:B_formula}	
	\mathcal{B} = zy' - \frac{1}{6}(1+3w)\lambda z^2.
\end{equation}
 The calculation of $\mathcal{E}$ is  more complicated. We start by using
the
distribution function (\ref{eq:MB_iso}) in order to calculate 
kinetic
energy $ K \equiv \frac{1}{2}m\int f(r,\upsilon)\upsilon^2 d^6\tau$, which
using the  dimensionless variables (\ref{eq:variables}) leads finally to the
dimensionless kinetic energy
\begin{equation}
\label{Kdimless}
	  \mathcal{K} \equiv \frac{KR}{GM^2} =
\frac{3}{2\mathcal{B}}\left[ 1 +
\frac{1}{6}(1+w)\frac{\lambda z^2}{\mathcal{B}}\right].
\end{equation}
Similarly, using (\ref{eq:phiTOT}) we define a dimensionless
expression for $\phi(0)$:
\begin{equation}
\label{dimlessphi0}
	\frac{m}{kT}\phi(0)  
	= - \left[\left( 1+ \frac{1+w}{2}\lambda\right) \int_0^z x
e^{-y}dx\right] + \frac{1+w}{4}\lambda z^2,
\end{equation}
and then inserting (\ref{dimlessphi0}) into
(\ref{eq:PhiNef}),(\ref{eq:PhiLef}),(\ref{eq:U_eff}),
 we calculate the dimensionless potential
energy as
\begin{align}
\label{eq:U_ND}
 \mathcal{U}\equiv\frac{UR}{GM^2} =&\frac{1}{2\mathcal{B}^2z} \left( 1 +
\frac{1+w}{2}\lambda\right) \int_0^z x^2y e^{-y} dx 
\nonumber \\ 
	& -\frac{1}{2\mathcal{B}} \left[ 1 + \frac{1}{6\mathcal{B}}(1+w)
\lambda z^2\right]
\nonumber \\
	& \times \left( 1 + \frac{1+w}{2}\lambda\right)\int_0^z x e^{-y} dx 
\nonumber \\
	 & + \frac{1}{12\mathcal{B}^2z}w\lambda \left( 1 +
\frac{1+w}{2}\lambda\right)
\int_0^z x^4 e^{-y} dx.
\end{align}
Finally, using (\ref{Kdimless}) and (\ref{eq:U_ND}), the dimensionless energy
is written as 
\begin{equation}
   \mathcal{E}= \mathcal{K}+\mathcal{U}.
\label{eq:E_ND}
\end{equation}

We note that the virial equation is modified both due to dark energy and the external pressure $P$. It is relatively easy to show that the virial equation becomes: 
\begin{equation}\label{eq:virial}
	2K + U_{N(\mbox{\footnotesize eff})} - 2U_{X(\mbox{\footnotesize eff})} = 3PV, 
\end{equation}
where $U_{N(\mbox{\footnotesize eff})}$ and $U_{X(\mbox{\footnotesize eff})}$ are the potential energy of the effective matter (that includes the dark energy particles) and the remaining dark energy potential energy, respectively. These two potential energies are the components of equation (\ref{eq:U_eff}). We verified numerically that the expressions (\ref{Kdimless}) and (\ref{eq:U_ND}) indeed satisfy the generalized virial equation (\ref{eq:virial}). Thus, the entropy extrema correspond to virialized configurations. 

One would naturally expect the relaxation process towards virialization, described by the Layzer-Irvine equation \cite{Peebles:1993} in an expanding Universe, to be affected, as well \cite{Basilakos:2010rs}. But in the present work this does not affect our results, since we derive our conclusions only by the radii at which an instability sets in. However, a separate analysis on the relaxation process would not only be interesting by its own right, but could also help understand some of our conclusions, such as the steepening of clusters density profile, as is demonstrated in section \ref{sec:clusters}.

\begin{figure}[ht]
\begin{center}
	\includegraphics[scale=0.6]{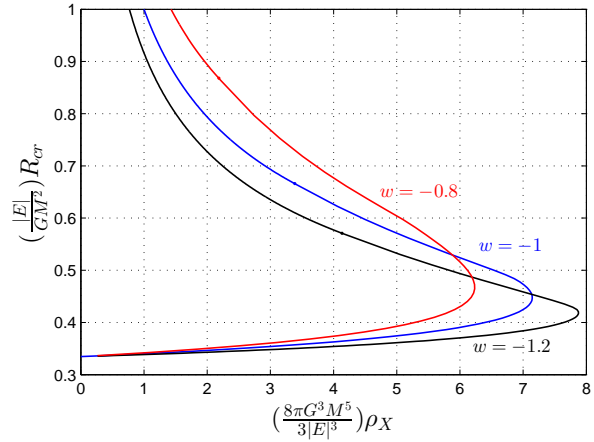}
	\caption{ The critical radius for fixed negative energy $E$ and mass
$M$ (microcanonical ensemble) versus
$\rho_X$ for cosmological constant ($w=-1$), quintessential
($w=-0.8$) and phantom ($w=-1.2$) dark energy. For each curve, the
instability domain
is inside it, that is no equilibria
exist between the two critical radii for some fixed
$\rho_X$. 
	\label{fig:RvsL_micro}}
\end{center} 
\end{figure}

\begin{figure}[ht]
\begin{center}
	\includegraphics[scale=0.6]{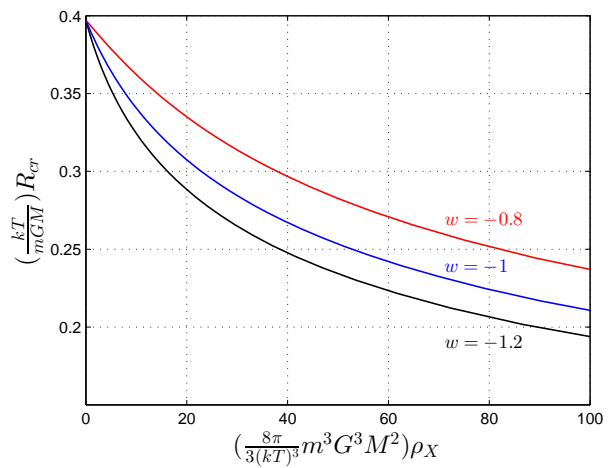}
	\caption{ The critical radius  for fixed temperature $T$ 
	and mass $M$ (canonical ensemble) versus $\rho_X$ for cosmological
constant ($w=-1$), quintessential
($w=-0.8$) and phantom ($w=-1.2$) dark energy. For each curve the instability
domain is the one below the
curve, and thus no equilibria
exist for radii smaller than the critical radius for some
fixed $\rho_X$.
	\label{fig:RvsL_can}}
\end{center} 
\end{figure}

Let us now focus on our scope of this section. That is to draw the critical radius, at which an instability sets in, 
in the microcanonical and canonical ensembles, with respect to $\rho_X$
for different $w$ values. To this end, we generate series of equilibria
solving
(\ref{eq:Emden}) with our code (that keeps $\mu$ constant) for different $\mu$
values
that correspond to different $\rho_X$. 
\begin{figure}[ht]
\begin{center}
	\includegraphics[scale=0.6]{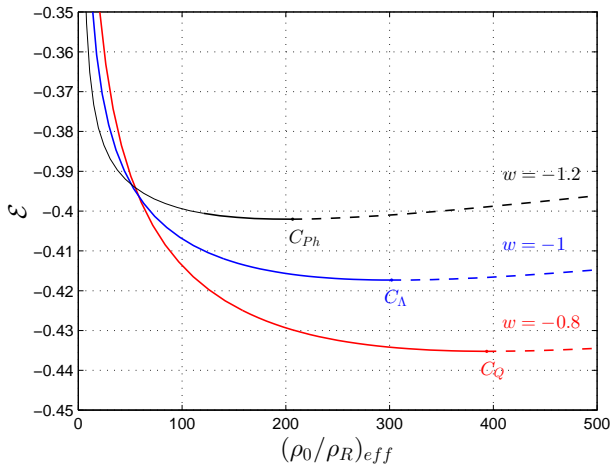}
	\caption{The series of equilibria expressed as $\mathcal{E} =
ER/GM^2$ versus
	the density contrast $(\rho_0/\rho_R)_{\mbox{\footnotesize eff}}$ for
a 
	simple cosmological constant $(w = -1)$, quintessential $(w=-0.8)$
and phantom $(w=-1.2)$
    dark energy. At points $C_{Ph}$, $C_\Lambda$ and $C_Q$ an instability
sets in. 
    The dashed curves correspond to unstable equilibria, while the solid
curves
    to stable equilibria.
	\label{fig:E_w}}
\end{center} 
\end{figure}
In Figure \ref{fig:E_w} we see the series of equilibria expressed by $\mathcal{E}$
w.r.t. the density contrast
$\rho_0/\rho_R$ for each isothermal sphere, for a specific $\mu$, 
for a simple cosmological constant $w = -1$, quintessential $w=-0.8$ and
phantom $w=-1.2$
dark energy. Each minimum is a turning point of stability in the
microcanonical ensemble.
On the other hand, the maximum in $\mathcal{B}$ is a turning point of
stability in the canonical ensemble.
We repeat the calculations of minima in $\mathcal{E}$ and maxima in
$\mathcal{B}$ 
for many $\mu$ values and finally plot these critical values in
Figure \ref{fig:RvsL_micro} 
and Figure \ref{fig:RvsL_can}. These critical values correspond to the desired
critical radii,
assuming $M$, $E$ constant in the microcanonical ensemble and $M$, $T$
constant
in the canonical ensemble.
 

In Figure \ref{fig:RvsL_micro}, for every $w$  we observe   the characteristic
reentrant behavior for the micorcanonical ensemble 
noticed firstly in \cite{Axenides:2012bf}.
For a fixed $\rho_X$, equilibria
do exist for arbitrarily small radii. Then, at some bigger radius an
instability sets in
and no equilibria are allowed. The system undergoes gravothermal catastrophe
\cite{Antonov:1962,Bell:1968}.
Note that recently, gravothermal catastrophe has been identified as similar
to Jeans instability \cite{Sormani:2013he}.
Returning to Figure \ref{fig:RvsL_micro}, we see that for some even bigger radius, 
let call it the ``reentrant radius'', the equilibria are restored. 
This effect is due to the dark energy, which introduces an harmonic,
repulsive force 
proportional to the radius. 
The equilibria above this radius have peculiar density profiles, with an
increasing
density towards the edge or with various local maxima
\cite{Axenides:2012bf,Axenides:2013hrq} . This means that if the walls
were absent, these states would correspond to perturbations that would follow
Universe's
expansion. Such perturbations would not collapse and could not lead to
structure formation.
Thus, \emph{the reentrant radius, defines the maximum
size of a perturbation that can lead to structure formation}. This resembles
exactly
the ``maximum turnaround radius'' noticed recently \cite{Pavlidou:2013zha}. 
We see in Figure \ref{fig:RvsL_micro} that the reentrant radius is smaller for
phantom dark energy than for quintessential one. Thus, large scale
structures are more easily formed in
a quintessential rather than a phantom universe. This effect is due to
the stronger repulsive pressure of phantom dark energy and it was
qualitatively expected, however in the above analysis it has been
incorporated quantitatively.

In Figure \ref{fig:RvsL_can} we notice that the  critical radius
in the canonical ensemble, is smaller for phantom 
rather than quintessential dark energy. Since the instability domain
is underneath the critical radius, we conclude again that quintessence
increases the instability domain, with respect to phantom dark energy and
the simple cosmological constant case. In the case of the canonical
ensemble,
this effect is only due to the effective particles that introduce additional
mutual attraction for quintessential dark energy. The negative pressure
is not important, because the instability sets in
for small radii where the repulsive force is irrelevant (it increases
with   distance).

Finally, we stress   that the minimum in $\mathcal{E}$
corresponds to some value of the density contrast
$(\rho_{0}/\rho_{R})_{\mbox{\footnotesize eff}}$, 
as can be seen in Figure \ref{fig:E_w}.
This is a critical value, where an instability sets in. For larger density
contrasts,  the equilibria are unstable and the system undergoes
gravothermal catastrophe. These unstable equilibria correspond to the dashed
curves
in Figure \ref{fig:E_w}.

\begin{figure}[ht]
\begin{center}
	\includegraphics[scale=0.6]{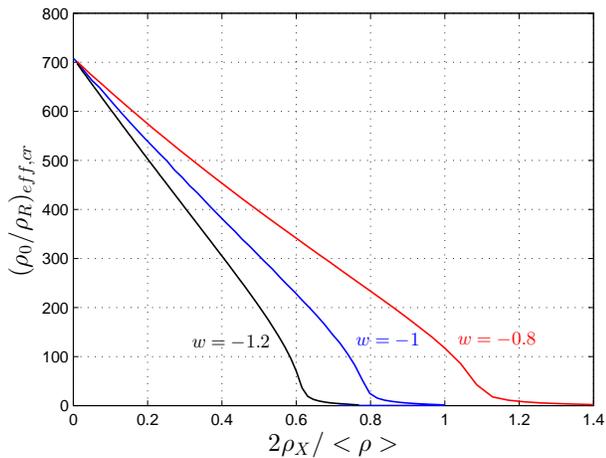}
	\caption{ The critical density contrast of effective density for
fixed radius $R$ and mass $M$ versus $\rho_X$ for cosmological constant
($w=-1$), quintessential ($w=-0.8$) and phantom ($w=-1.2$) dark energy
in the microcanonical ensemble. For
each curve, all equilibria above it are unstable, that is the ones with
density
contrast bigger than the critical value for some fixed $\rho_X$.
	\label{fig:dens_cont}}
\end{center} 
\end{figure}

In the next section we will use  this critical
density contrast, which for completeness is depicted
in
Figure \ref{fig:dens_cont}, where we see that the critical density contrast is
smaller for phantom rather than quintessential dark energy. Equilibria
corresponding to density contrast value that lies above each curve are
unstable, while the ones underneath each curve are stable.

\section{Galaxy Clusters in the presence of dark energy}\label{sec:clusters}

In the previous section we performed a detailed general analysis of the
stability of self-gravitating gas in the presence of a general dark-energy
component. In this section we will apply it in the  specific case of galaxy
clusters, which are dominated by the dark matter halo.
Our goal  is to first estimate the effect
of dark energy on
the density profile of the dark-matter halo and secondly to examine whether
any constrain
on the equation-of-state of dark energy can be deduced from observational
facts of galaxy clusters. We will perform the analysis in the most simple
set-up, assuming hydrostatic equilibrium, isothermal distribution and
spherical symmetry.

 Galaxy clusters are the largest, virialized, self-gravitating bound
systems in the Universe. They have been the focus of intense study for
several
decades, among other reasons  because they provide crucial information on the
formation of large-scale structure and on estimates of cosmological
parameters \cite{Allen:2011zs}. Galaxy clusters are consisted of three
components. Dark matter is the main component holding about $80-90\%$ of the
total mass, the X-ray emitting hot intracluster medium (ICM) makes up about
$10-20\%$ and only a small fraction $\sim 1\%$ corresponds to cold gas, dust
and stars found mainly in galaxies. The intracluster medium is hot plasma
with temperature about $2-10keV$, consisted mainly of ionized hydrogen and
helium and electrons. It emits X-rays due to thermal bremsstrahlung. All
three components are found to be approximately in hydrostatic equilibrium
\cite{Sarazin:1986rmp,Kravtsov:2012zs} inside the gravitational well of the
cluster, dominated by the dark matter halo. The hypothesis of hydrostatic
equilibrium 
corresponds to the assumption that self-gravity is halted only by thermal
pressure. Non-thermal pressure is found to contribute at about
$10\%$ \cite{Faltenbacher:2004cg, Schuecker:2004hw, Rasia:2006ra} to the
total pressure. Dark matter is assumed to be collisionless, although recently
the possibility to be slightly self-interacting is being inspected
\cite{Spergel:1999mh,Randall:2007ph,Bradac:2008eu,Chan:2013ysa}. The ICM is
nearly isothermal, at least apart from the central regions
\cite{Reiprich:2001zv, LaRoque:2006te}.

Due to the equivalence between inertial and gravitational mass, orbits
in a gravitational system are independent of the mass of the orbiting
particles. Therefore, it is legitimate to assume that different species in a
relaxed, spherically-symmetric, gravitational system have the same average
specific kinetic energy. In a gas system, equilibrium implies energy
equipartition between different species, while for a relaxed, gravitational
system the corresponding principle would be the common velocity dispersion,
because of the equivalence principle. Indeed, simulations of dark matter
haloes  \cite{Host:2008gi,Evrard:2007py} do  indicate that
$
	 (k\overline{T}/m)/\sigma^2_{DM} \simeq 1
$
while observational data indicate \cite{Xue:2000pp} that
$(k\overline{T}/m)/\sigma^2_{gal} \simeq 1
$,
where $\overline{T}$ is the mean temperature of the ICM, $\sigma^2_{DM}$ is
the dark matter velocity dispersion, $\sigma^2_{gal}$ is the galaxy velocity
dispersion, $m \simeq 0.6m_p$ is the mean particle mass of ICM and $m_p$ is
the proton mass. Therefore, we have strong arguments to justify the
consideration 
\begin{equation}\label{eq:sigmaT}
	\sigma^2_{DM} = \sigma^2_{gal} = \frac{k T}{m},
\end{equation}
where $T$ is the temperature of ICM in an isothermal distribution. 
Consequently, under these assumptions, the three components of a galaxy
cluster have the same density distributions,
leading to the total distribution:
\begin{equation}
\label{eq:dens_gc}
	\rho(r) = \rho_0 e^{-\frac{m}{kT}[\phi(r)-\phi(0)]}.
\end{equation}

\begin{table}[ht]
\begin{center}
	\begin{tabular}{l | l l l l l}
	Cluster & $z$ &  $M_{vir}$ & $R_{vir}$ & $T$ &
$\frac{2\rho_X}{\langle\rho\rangle}$ \\
	\hline
MS 0906+11	& 0.1704	& 8.3		& 1737		& 6.1	&
0.0103
\\
MS 1224+20	& 0.3255	& 3.8		& 1226		& 4.8	&
0.0079
\\	
MS 1358+62	& 0.3290	& 10.3		& 1706		& 6.7	&
0.0079
\\
MS 1512+36	& 0.3727	& 3.3		& 1139		& 4.1	&
0.0073
\\
MS 1621+26	& 0.4275	& 12.3		& 1715		& 8.1	&
0.0067
\\
A68			& 0.2550	& 10.5		& 1790		& 8.0
& 0.0089	
\\
A267		& 0.2300	& 7.5		& 1623		& 5.9	&
0.0093
\\
A963		& 0.2060	& 6.5		& 1569		& 6.6	&
0.0097
\\
A1763		& 0.2230	& 13.5		& 1982		& 7.7	&
0.0094
\\
A2218		& 0.1756	& 8.8		& 1766		& 7.0	&
0.0102
\\
A2219		& 0.2256	& 11.3		& 1865		& 9.8	&
0.0094
\\
\hline
	\end{tabular}
	\caption{ The redshift $z$, virial mass
$M_{vir}(10^{14}h^{-1}M_\odot)$, virial radius $R_{vir}(h^{-1}kpc)$,
temperature $T(keV)$ and the
ratio of the cosmological constant to the mean density for $\rho_X =
6.5\cdot 10^{-30}gr/cm^3$ for some galaxy clusters. The virial mass is
calculated by Hoekstra \cite{Hoekstra:2007nc} with a Navarro-Frenk-White (NFW) fit to weak-lensing
data.
	\label{tab:clusters}}
\end{center} 
\end{table}

In realistic situations not all
components have the same distributions, but since in this work we are
interested in estimating the effect of dark energy to the
distribution of the cluster, and not determining the exact
distributions, we expect that deviations from expressions
(\ref{eq:sigmaT}) and (\ref{eq:dens_gc}) would not alter the dark-energy
effect. Since clusters are dominated by the dark matter halo, our results
  hold for the halo's profile. Additionally, it is interesting to
note that  the above assumptions
are exactly the same with those of the so-called  truncated isothermal
sphere (TIS) model \cite{Shapiro:1998zp,Iliev:2001he}. In TIS model
 the dark-matter halo is assumed to be spherical, isothermal and in
equilibrium (that is an isothermal sphere), formed from the collapse and
virialization of ``top-hat'' density-perturbations. The TIS scenario is a
unique,
non-singular solution of the Emden equation, modified with a cosmological
constant \cite{Iliev:2001he}, corresponding to the minimum-energy solution
under constant external pressure, while the gas and the dark-matter halo are
assumed to have the same distribution as in equation
(\ref{eq:dens_gc}). The difference of TIS with our analysis is that we 
consider all non-singular solutions of the
modified Emden equation (\ref{eq:Emden}) and not only the specific TIS one.
This model is in a very good agreement   
with simulations and observations \cite{Shapiro:2004vp}, at least outside the
inner regions of the cluster. In the inner regions the TIS model
predicts a soft core, however collisionless $N$-body simulations of dark
matter haloes predict a cusp, rather than a core.
The main reason
for this difference is the assumption of isothermality: the $N$-body
simulations are
  nearly isothermal apart from a small region dip near the center,
which causes the cuspy profile, while  the self-interacting dark
matter is possible to form a central core \cite{Ahn:2004xt}. Moreover, in
contrast with simulations, 
observations at small scales favor the existence of a central core in dark
matter haloes, 
a problem called the ``core-cusp problem'' \cite{Hui:2001wy,deBlok:2009sp}. 
Regarding our work, the agreement of the non-singular, isothermal sphere 
with observations and simulations, besides the inner region, is sufficient
for our purposes.

Let us now proceed to the estimation of the effect of dark
energy to the density profile of galaxy clusters. Note that the
effective mass and density profile are the
ones measured by indirect measurements such as gravitational lensing and 
hydrostatic equilibrium. Thus, in all of our subsequent analysis of galaxy
clusters we use the effective values of density contrast (for the
cosmological constant $w=-1$ however the effective density is identical to
the matter density). 

In the estimation procedure we will need the quantity
$2\rho_X/\langle\rho\rangle$ in galaxy clusters, where  
$\langle\rho\rangle = 3M_{vir}/4\pi R^3_{vir}$ is the average density of the
cluster as it arises from observations. In order
to extract this  observation-related value, we work with a
sample of $11$ clusters taken by Hoekstra \cite{Hoekstra:2007nc}, who
performed a model-independent
analysis based on weak-lensing measurements. We used
only these clusters that have such virial radius, virial mass and temperature
that can correspond to an isothermal equilibrium.
The virial radius, the virial
mass, the temperature and the ratio $2\rho_X/\langle\rho\rangle$ for each
cluster is shown in
Table \ref{tab:clusters}. We have used $\rho_X \simeq
6.5\cdot 10^{-30}gr/cm^3$ \cite{Hinshaw:2012aka}. 
The mean ratio $2\rho_X/\langle\rho\rangle$ is:
\begin{equation}
\label{eq:MND_value}
	\frac{1}{\mu} = \langle\frac{2\rho_X}{\langle\rho\rangle}\rangle =
0.009,
\end{equation}
and the mean $\mathcal{B} = G M_{vir} m/kT  R_{vir}$ is
\begin{equation}\label{eq:GMB_value}
	\langle \mathcal{B} \rangle = 2.07.
\end{equation}

\begin{figure}[ht]
\begin{center}
	\includegraphics[scale=0.6]{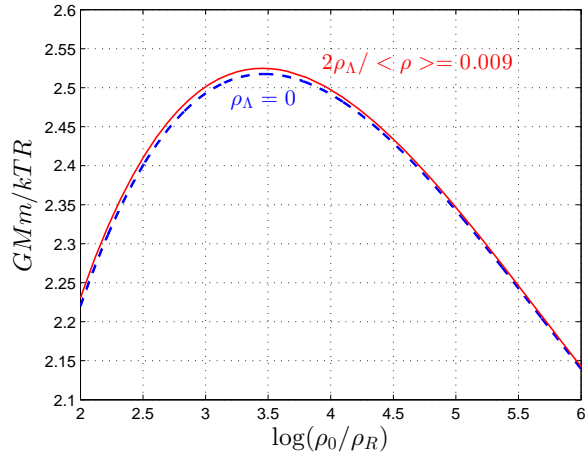}
	\caption{ The series of equilibria without dark energy (dashed blue
line) and with a cosmological constant, i.e. $w=-1$, (solid red line) for
$2\rho_\Lambda/\langle\rho\rangle = 0.009$ that corresponds to our galaxy
clusters
sample,
expressed as the ratio $GMm/kTR$ versus the natural logarithm of the central
to edge density ratio of the cluster. At $GMm/kTR = 2.07$ correspond two
density profiles at each case ($\rho_\Lambda = 0$ and $\rho_\Lambda \neq
0$). 
The profiles with the higher density contrast (after the peak)
are the ones relevant to galaxy clusters. The case $\rho_\Lambda \neq 0$
has greater density contrast than the case $\rho_\Lambda = 0$.
	\label{fig:GMmbR}}
\end{center} 
\end{figure}
\begin{figure}[ht]
\begin{center}
	\includegraphics[scale=0.6]{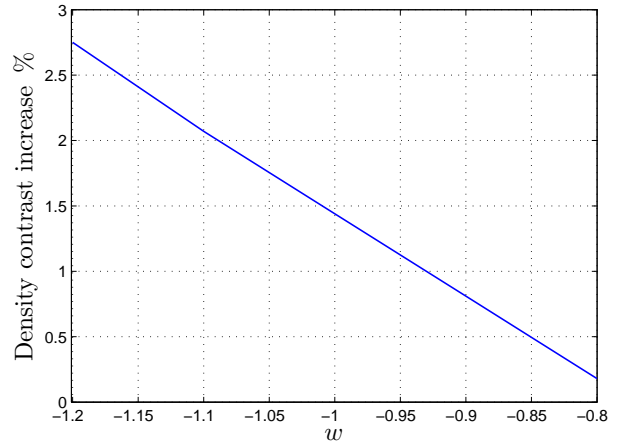}
	\caption{ The effect of dark energy to the density profile of galaxy
clusters. The percentage increase of the effective density contrast
$(\rho_0/\rho_R)_{\mbox{\footnotesize eff}}$, i.e. the center divided by the
edge density, for
$\rho_X = 6.5\cdot 10^{-30}gr/cm^3$ with respect to the case
$\rho_X = 0$ versus $w$, for fixed temperature, total mass and radius.
	\label{fig:percent}}
\end{center} 
\end{figure}
Using the computer code of the previous section, and for dimensionless mass
(equation (\ref{eq:M_dless})) $\mu = 111.11$ that corresponds to
(\ref{eq:MND_value}), we calculate numerically the corresponding value to
(\ref{eq:GMB_value}) of the effective density contrast
$(\rho_0/\rho_R)_{\mbox{\footnotesize eff}}$ both in the presence of dark
energy and for $\rho_X = 0$ and find how much is the profile altered by the
presence of dark energy. For fixed radius, mass and temperature, we find the
counter-intuitive result that dark energy, either quintessential or phantom
or a cosmological constant, tends to steepen the density profile increasing
the density contrast. This can be seen in figure \ref{fig:GMmbR} for density
contrast values greater than the peak, which are the ones relevant to
observations. We believe the reason is that for fixed temperature, mass and
radius, the extra outward pressure introduced by dark energy, enables more
mass to be concentrated towards the center of the cluster. For the case
$w=-1$, i.e.
 the cosmological constant, we find that the density profile is steepened by
an amount of $1.5\%$.
Finally, we mention that  the effect of dark energy
becomes stronger as  $w$ becomes more negative, as can be seen in
figure \ref{fig:percent}.

Finally, let us propose a method to
impose constraints on the dark-energy equation-of-state parameter. As we
discussed in   section \ref{sec:instab},  as $w$ attains more
negative values, the critical density contrast (the maximum allowed before
gravothermal catastrophe occurs) decreases, as in figure \ref{fig:dens_cont}.
Therefore, we can use this property in order to find a minimum allowed value
of $w$. If the minimum possible density contrast, corresponding to an
isothermal distribution
can be determined from clusters' observations, then the critical density
contrast
can be no smaller than this value. Hence, provided the minimum observational
density contrast is given, 
we can numerically calculate the $w$ which has
critical density contrast equal to this value. This $w$ value would be
 the minimum possible in order for the cluster to be in equilibrium.

\section{Conclusions}\label{Conclusions}

In this work  we studied the effect of dark energy on the stability of
isothermal spheres  for various values of $\rho_X$ (in section
\ref{sec:instab}), and furthermore, based on this
analysis,  we focused on the effect of dark energy on
galaxy clusters (in section \ref{sec:clusters}).

\indent We assumed a linear and constant equation of state for dark energy
and we investigated the effect on self-gravitating gas bound by external
pressure (walls), in the Newtonian limit. Dark energy introduces a repulsive
force due to the negative pressure, generated by an effective potential
(equation (\ref{eq:PhiLef}),
but it additionally  introduces   ``dark energy'' particles through an
effective
density, given in equation (\ref{eq:rhoeff_def}).
These dark energy particles strengthen attraction in case of quintessential
dark energy ($w>-1$) and weaken attraction in case of phantom dark energy ($w
< -1$). The total effect, however, of repulsive potential and dark energy
particles, is in all cases repulsive. We calculated the entropy extremum and
we
found that it corresponds to a Boltzmann distribution 
for the effective density (see equations (\ref{eq:rhoeff_dis}) and
(\ref{eq:rho_DIS})).

Then, we focused on the effect of dark energy in the stability of
isothermal spheres. This effect can be summarized in
figures \ref{fig:RvsL_micro}, \ref{fig:RvsL_can} and \ref{fig:dens_cont}.
The microcanonical ensemble
(fixed energy) in the presence of a cosmological constant is known to present
a reentrant phase transition \cite{Axenides:2012bf}, that is for some fixed
$\rho_X$ there exist two critical radii and no equilibria exist between
these two values. The upper radius is called the reentrant radius
and at this radius equilibria are restored. This equilibria have
increasing density towards the edge and correspond to perturbations
that follow the expansion in an expanding Universe.
Thus, \emph{the reentrant radius, defines the maximum
size of a perturbation that can lead to structure formation}.
Quintessence increases the reentrant radius,
while phantom dark energy decreases it, as can be seen in
Figure \ref{fig:RvsL_micro}. Therefore, a quintessence universe
is expected to present richer large scale
structure, with more and larger bounded systems, than a phantom universe. 
In the canonical ensemble (fixed temperature) there is
only one critical radius less than which there are no equilibria. 
Quintessence increases this critical radius with respect to
the simple cosmological constant case, while phantom dark
energy decreases it, as can be seen in Figure \ref{fig:RvsL_can}. 
Thus, quintessential dark energy enlarges the instability domain, while
phantom
dark energy narrows it, with respect to the cosmological
constant.
Finally, we inspected how the critical effective density contrast (the
critical central
to edge effective density ratio), corresponding to 
gravothermal catastrophe (fixed energy) is affected by dark energy. The
result is shown in Figure \ref{fig:dens_cont}. Quintessential dark energy increases
the critical density contrast, while phantom dark energy decreases it. This
implies that in a quintessence universe more condensed large scale structures
are formed.

Regarding the second part of this work (section \ref{sec:clusters}),
let us remark that we find a rather counter-intuitive result. Dark energy
causes the density profile of galaxy clusters to be more centrally
concentrated. That is, for fixed mass, radius and temperature, the system
will
equilibrate in a larger density contrast (central to edge ratio
$\rho_0/\rho_R$) in the presence of dark energy. This is manifested in
Figure \ref{fig:GMmbR}. It seems as if for
these equilibria, the extra outward pointing pressure of dark energy is added
to the thermal pressure, enabling
the system to equilibrate in a more condensed state. This might be 
associated with the fact that these equilibria are
unstable under variations that preserve the temperature, namely isothermal
collapse. This is evident in Figure \ref{fig:GMmbR}, where the equilibria 
corresponding to galaxy clusters are the ones after the peak and hence
are unstable under isothermal collapse. 
However, they are stable under variations that preserve the energy
(instability in this case would correspond to gravothermal catastrophe), at
least up to some greater density contrast value. Most importantly, this
effect, that is the steepening of the density profile due to dark energy,
corresponds in the case $w=-1$ to equilibration of the cluster for the same
$M$,
$T$ and $R$ at about $1.5\%$ greater density contrast. 
The effect is getting stronger as $w$
attains more negative values. This is evident in Figure \ref{fig:percent}. 
We note that clustered dark energy is found in Ref. \cite{Basilakos:2010rs}
to produce even more concentrated structures than the homogeneous vacuum energy,
considered in the current work.
The case of clustered dark energy is not considered here, since
it would introduce various complexities in the analysis, such as the
modification in the virial theorem. We think the relevance
of the current formulation with Ref. \cite{Basilakos:2010rs}
should be explored further. 

Last but not least, we proposed a method to constrain phantom dark
energy from galaxy clusters observations. As we have seen in section
\ref{sec:instab} the critical density contrast at which the instability
sets in is decreasing with decreasing $w$. Therefore, if one can
determine, based on galaxy clusters observations, the density contrast
corresponding to static isothermal equilibrium, then one can determine the
minimum $w$ as the one that has critical density contrast equal to this
value. 

We close by making a comment on the generality of our results. In the above
analysis we considered only   
the case of a linear and constant equation of  state, 
in order to understand the basic effects of dark energy. Clearly, a
divergence from these assumptions deserves separate investigation, since the
results could quantitatively (or even qualitatively) change. Such could be
the more general case of a time-varying $w$ and/or time-varying
cosmological constant
\cite{Shapiro:2000dz,Polyakov:2009nq,Basilakos:2010rs,Sola:2013gha,
Basilakos:2013xpa}, remaining in the linear equation of state, or even going
to more general equation of states such is the generalized polytropic one
\cite{Chavanis:2013gds,Chavanis:2012pd,Chavanis:2012kla,Chavanis:2012uq}. 
These extensions are under investigation and are going to be
presented in a future publication. Finally, note that in reality, the dark
energy sector may have an effective nature, and thus its equation of state
too, not corresponding to fundamental fields or degrees of freedom
\cite{Sola:2005et,Sola:2005nh,Basilakos:2013vya}. In this case its implication on the
galaxy cluster might change too, and thus it might offer a way to
distinguish amongst the various dark energy scenarios. 

\begin{acknowledgments}
The research
of ENS is implemented
within the framework of the Action ``Supporting Postdoctoral Researchers''
of the Operational Program ``Education and Lifelong Learning'' (Actions
Beneficiary: General Secretariat for Research and Technology), and is
co-financed by the European Social Fund (ESF) and the Greek State.
The work of M.A.  was supported in part by the
General Secretariat for Research and Technology of Greece and
the European Regional Development Fund
MIS-448332-ORASY (NSRF 2007-13 ACTION, KRIPIS).
\end{acknowledgments}
\appendix

\appendix
\section{}\label{app:A}

We derive the TOV equations (\ref{eq:TOV1}) and (\ref{eq:mprime}). Any
spherically symmetric metric can be written in the form:
\[
	ds^2 = e^\nu c^2 dt^2 - e^\lambda dr^2 - d\Omega,
\]
where in general $\nu = \nu(r,t)$ and $\lambda = \lambda(r,t)$. 
The Einstein's equations
\[
	R^\mu_\nu - \frac{1}{2} R \delta^\mu_\nu = \frac{8 \pi G}{c^4}
T^\mu_\nu
\]
give:
\begin{align}
\label{eq:tensorcom1}	
& \frac{8\pi G}{c^4}T^0_0 = e^{-\lambda}\left( \frac{\lambda '}{r} -
\frac{1}{r^2}\right) + \frac{1}{r^2} 
\\
\label{eq:tensorcom2}	
& \frac{8\pi G}{c^4}T^1_1 = -e^{-\lambda}\left( \frac{\nu '}{r} +
\frac{1}{r^2}\right) + \frac{1}{r^2} 
\\
\nonumber
& 
 \frac{8\pi G}{c^4}T^2_2 = \frac{8\pi G}{c^4}T^3_3 =
		 -e^{-\lambda}\left( \frac{\nu ''}{2} -
\frac{\lambda'\nu'}{4} + \frac{{\nu '}^2}{4} +
\frac{\nu'-\lambda'}{2r}\right) 
\\
\label{eq:tensorcom3}	
& \quad\quad\quad\quad\quad\quad\quad\quad\quad  
+ e^{-\nu}\left( \frac{\ddot{\lambda}}{2} + \frac{\dot{\lambda}^2}{4} -
\frac{\dot{\lambda}\dot{\nu}}{4}\right) 
\\
\label{eq:tensorcom4}	
& \frac{8\pi G}{c^4}T^1_0 = -e^{-\lambda} \frac{\dot{\lambda}}{r} 
\\
\label{eq:tensorcom5}		
& \frac{8\pi G}{c^4}T^0_1 = e^{-\nu} \frac{\dot{\lambda}}{r},
\end{align}
where a prime denote differentiation w.r.t. $r$ and a dot w.r.t. $t$. We set
the
energy momentum tensor to be that of a perfect fluid in the presence of
$\rho_X$ with $p_X = w \rho_X c^2$, that is
\begin{equation}\label{eq:EMtensor}
	T^\mu_\nu = (\tilde{p} + \tilde{\rho} c^2)g_{\alpha \nu}
\frac{dx^\mu}{ds}\frac{dx^\alpha}{ds} - \tilde{p} \delta^\mu_\nu,
\end{equation}
with $\tilde{\rho} = \rho + \rho_X$ and $\tilde{p} = p + w\rho_X c^2$.
At the equilibrium it is just $T^\mu_\nu \rightarrow (\tilde{\rho}c^2, -
\tilde{p},- \tilde{p},- \tilde{p})$ and $\dot{\lambda} = 0$, $\dot{\nu} = 0$.
Substituting into Einstein's equations and after some calculations,
equations (\ref{eq:tensorcom1}-\ref{eq:tensorcom5}) give:
\begin{align}
\label{eq:tolequi1}	
& \frac{dp}{dr} = -\frac{1}{2}(p + \rho c^2 +
(1+w)\rho_X c^2) \nu' \\
\label{eq:tolequi2}	
& \frac{8\pi G}{c^2}\rho = e^{-\lambda}\left(
\frac{\lambda '}{r} - \frac{1}{r^2}\right) + \frac{1}{r^2} 
						- \frac{8\pi
G}{c^2}\rho_X \\
\label{eq:tolequi3}	
& \frac{8\pi G}{c^4} p  = e^{-\lambda}\left( \frac{\nu
'}{r} + \frac{1}{r^2}\right) - \frac{1}{r^2} 
						- w\frac{8\pi
G}{c^2}\rho_X ,
\end{align}
which by the transformation
\begin{equation}\label{eq:lm}
	e^{-\lambda} = 1 - \frac{2G\mathcal{M}(r)}{r c^2} - \frac{8\pi G}{3
c^2}\rho_X r^2
\end{equation}
become just two equations, namely
\begin{widetext}
\begin{align}
& \frac{dp}{dr} = -\left[\frac{p}{c^2} + \rho + (1+w)\rho_X\right]\left[
\frac{G\mathcal{M}(r)}{r^2} + 4\pi G \frac{p}{c^2}r + \frac{4\pi
G}{3}\rho_X r(1+3w)\right]
 \left( 1 - \frac{2G\mathcal{M}(r)}{rc^2} - \frac{8\pi
G}{3c^2}\rho_X r^2\right)^{-1}		\\
& \frac{d\mathcal{M}(r)}{dr} = 4\pi\rho r^2.
\end{align}
\end{widetext}


\bibliography{DM11404_FINAL}
\bibliographystyle{h-physrev}

\end{document}